\documentclass[%
 reprint,
superscriptaddress,
 amsmath,amssymb,
 aps,
 twocols
]{revtex4-1}

\usepackage{verbatim}
\usepackage{graphicx}% Include figure files
\usepackage{dcolumn}% Align table columns on decimal point
\usepackage{bm}
\usepackage{mathtools}                                %%%
\DeclarePairedDelimiter\abs{\lvert}{\rvert}           %%%
\usepackage{float}                                    %%%
\newcommand\ch[1]{\ensuremath{\mathcal{H}_{#1}}}      %%%
\usepackage{graphicx}                                 %%%
\usepackage{subcaption}                               %%%
\usepackage{hyperref}                                 %%%
\usepackage[usenames, dvipsnames]{color}
\usepackage{color}
\begin{document}

\preprint{APS/123-QED}

\title{A deep learning approach to the structural analysis of proteins}% Force line breaks with \\

\author{Marco Giulini}
\author{Raffaello Potestio}%
 \email{raffaello.potestio@unitn.it}
\affiliation{Physics Department, University of Trento, via Sommarive, 14 I-38123 Trento, Italy}
\affiliation{INFN-TIFPA, Trento Institute for Fundamental Physics and Applications, I-38123 Trento, Italy}


%\affiliation{
%\affiliation{
%\affiliation{%

\date{\today}% It is always \today, today,

\begin{abstract}
Deep Learning (DL) algorithms hold great promise for applications in the field of computational biophysics. In fact, the vast amount of available molecular structures, as well as their notable complexity, constitutes an ideal context in which DL-based approaches can be profitably employed.
To express the full potential of these techniques, though, it is a prerequisite to express the information contained in the molecule's atomic positions and distances in a set of input quantities that the network can process. Many of the molecular descriptors devised insofar are effective and manageable for relatively small structures, but become complex and cumbersome for larger ones. Furthermore, most of them are defined locally, a feature that could represent a limit for those applications where global properties are of interest.
Here, we build a deep learning architecture capable of predicting non-trivial and intrinsically global quantities, that is, the eigenvalues of a protein's lowest-energy fluctuation modes. This application represents a first, relatively simple test bed for the development of a neural network approach to the quantitative analysis of protein structures, and demonstrates unexpected use in the identification of mechanically relevant regions of the molecule.
%\begin{description}
%\end{description}
\end{abstract}

\pacs{Valid PACS appear here}% PACS, the Physics and Astronomy
\maketitle

%\tableofcontents

\section{Introduction}

Proteins are the most versatile biological molecules, as they cover roles ranging from ``mere'' structural support (e.g. in the cytoskeleton) to active cargo transport, passing through enzymatic chemistry, protein folding chaperoning, communication, photochemical sensing etc. The impressive variety of activities, sizes, shapes and functions proteins show is largely due to the LEGO-like capacity of the polypeptide chain, as well as to the polymorphic chemistry entailed in the 20 amino acids they are made of. According to the well-established central dogma of biology, the amino acid sequence of the protein dictates its three-dimensional structure, which in turn determines and enables the molecule's function. It should thus come as no surprise that protein structures have been thoroughly investigated at all levels, from the fundamental, experimental determination of the arrangement of their atoms in space (e.g. by means of X-ray crystallography or nuclear magnetic resonance) to computer-aided analyses aimed at understanding the interplay between sequence, structure, and function. These latter studies are carried out through {\it in silico} representations of the molecules whose resolution ranges from atomistic --as it is typically the case in molecular dynamics (MD) \cite{md_general_method, md_sim_biomol}-- to simplified, {\it coarse-grained} models \cite{Takada2012,Noid2013,Saunders2013,Potestio2014}, where several atoms are lumped together in sites interacting {\it via} effective potentials. Furthermore, the field of protein bioinformatics has boomed in the past few decades, where the wealth of available sequences and structures has been exploited to investigate structure prediction, protein-protein interaction, docking, protein-related genomics etc. (see e.g. \cite{pazos2016,wu2017} for recent, comprehensive reviews). 

The availability of a large number of instances of the protein space (be that sequence or structure) and the necessity to perform dataset-wide analyses and screening of their properties naturally leads one to wonder whether one could take advantage of the recent progresses achieved by machine learning approaches, in particular deep learning (DL). The latter is a subset of the wide class of machine learning computational methods, and has been successfully applied to a fairly wide spectrum of areas of science \cite{nat-bengio,bengio}, ranging from neuroscience \cite{marblestone_2016} to image and speech recognition \cite{review_dl_image_recognition, imagenet}. In the field of Computational Chemistry much effort has been devoted to the identification of the variables that are able to provide a comprehensive description of a chemical compound (\textit{molecular descriptors}). These features are usually designed in order to be applied to elements of the Chemical Compound Space (CCS), the theoretical combinatorial set of all possible compounds that could be isolated and constructed from all combinations and configurations of atoms \cite{ccs_2004, ccs_2012, ccs_2015}. Several examples of descriptors are present in the literature \cite{todeschini, ecf, ecf2, mold2, moldescr_vracko,imbalzanoJCP2018,grisafiACSCS2018}: they proved to be extremely useful in the development of predictive models about a huge variety of molecular properties. Nevertheless, the size of the CCS is limited from above by the Lipinski rules \cite{bohacek, lipinski}, that set the maximum molecular weight to 500 atomic mass units. It thus appears evident that the vast majority of structures studied in biophysics, such as proteins, nucleic acids, and polysaccharides, falls well beyond this value. As an example, the structures conserved in the Protein Data Bank (PDB) have a molecular weight ranging from few hundreds to hundred thousand daltons. One of the biggest issues in the application of DL-based approaches to biophysical problems thus consists in defining a flexible and robust method to properly encode the huge amount of information contained in these structures ({\it feature extraction}).

Nonetheless, deep learning algorithms are enjoying increasing popularity in the context of biological and condensed matter physics as well. In particular, Behler and Parrinello employed DL to construct accurate potential energy surfaces at low computational cost \cite{bp, behler}, while Dellago {\it et al.} \cite{dellago} built DL architectures capable of identifying local phases in liquids. More recently, Feinberg {\it et al.} \cite{feinberg} developed PotentialNet, a DL-based model for predicting molecular properties. At the same time, Wehmeyer and No\'e \cite{wehmeyer_noe} have implemented a complex DL algorithm (time-lagged autoencoder) able to perform efficient dimensionality reduction on Molecular Dynamics trajectories. Notably, promising works by Peter {\it et al.} \cite{cpeter} and Zhang {\it et al.} \cite{deepcg} extended the use of such algorithms to the field of coarse-grained models.

In spite of the recent encouraging attempts, a straightforward approach to a DL-based structural analysis protocol for the study of large macromolecules is lacking. In the present work, we aim at moving a step forward in this direction through the construction of a DL model to analyze protein structures. We set ourselves a relatively simple goal, that is, to predict the ten lowest eigenvalues of an exactly solvable coarse-grained model of a protein's collective fluctuations. The first aim of this work consists in identifying the procedure of feature extraction that is most suitable to our task. Second, we show that the application of a simple, standard and computationally not expensive DL architecture to the selected features gives satisfactory results, suggesting that more complex tasks will be attainable with similar, more refined networks. It is worth pointing out that although the development of a DL-based predictive model leads to a significant computational gain with respect to the exact algorithmic procedure, this is not the purpose of this work: here, we focus on demonstrating the viability of a DL-based approach to a specific class of problems in computational biophysics.

\section{Materials and Methods}

In this section we first summarise a few relevant concepts about Deep Neural Networks (DNNs), and specifically on Convolutional Neural Networks (CNNs). Subsequently, we provide a brief overview of the protein models of interest for our work, that is, Elastic Network Models (ENMs).

The raw data employed in the present work, including PDB files, protein structures datasets, CNN training protocols, trained networks and related material are publicly available on the ERC VARIAMOLS project website {\tt http://variamols.physics.unitn.eu} in the Research output section.

\subsection{Convolutional neural network model}

Born in the Fifties as theoretical, simplified models of neural structure and activity, neural networks are becoming an increasingly pervasive instrument for the most diverse types of computation. In particular, the tasks in which DNNs excel are those that can be reduced to classification, pattern recognition, feature extraction and, more recently, even a rudimentary (yet impressive) creative process. DNNs can be understood as a very complex form of {\it fitting procedure}, in that the parameters of the network are set through a process of training over a large dataset of items for which the outcome value is known; a prototypical example is that of a network endowed with the task of distinguishing images of dogs from those of cats, which is ``trained'' by proposing to it several images of the two types and changing the parameters so that the outcome label corresponds to the correct one. The following step is the validation of the network's effectiveness onto a complementary dataset of input instances that have not been employed in the training. The predictive power of DNNs is largely due to the non-linear character of the functions employed to connect one ``neuron'' to the following. This characteristic makes them substantially more flexible and versatile than multi-dimensional linear regression models, albeit also more obscure to comprehend in their functioning.

Deep Feedforward Networks (also called Multilayer Perceptrons (MLPs) or Artificial Neural Networks (ANNs)) are the most known class of Machine Learning algorithms \cite{bengio}. Given some input values $\bm{x}$ and an output label $y$ (categorical or numerical), MLPs assume the existence of a stochastic function $F$ of $x$  such that $y = F(\bm{x})$. They define a mapping $y =  f(\bm{x}; \bm{W})$ and try to learn the value of parameters $\bm{W}$ that give the best approximation of $F$. This function $f(\bm{x}; \bm{W})$ is the composition of $n$ different functions (usually called \emph{layers}), where $n$ is the depth of the MLP.

\begin{equation}\label{eqmlp}
f(\bm{x}) = f_{n}(f_{n-1}(...f_{2}(f_{1}(\bm{x}))))
\end{equation}

The function $f_{1}$, which directly acts on the input data, is called the \textit{input layer}, while $f_{n}$ is the \textit{output layer}. $f_{2} \cdots f_{n-1}$ and, more generally, all the intermediate layers are called \textit{hidden} because their scope is to translate the results coming from the first layer into an input that can be processed by the output layer. Most importantly, the functions $f_n$, also called activation functions, are {\it non-linear}: in fact, a neural network with only linear activations in the hidden layers would be equivalent to a linear regression model \cite{bengio}.

Eq. \ref{eqmlp} shows that a layer can be thought of as a function that takes  a vector as input and gives  a different vector as output. One can also imagine a layer as a set of vector-to-scalar functions (neurons) that act in parallel \cite{bengio}. Neurons are the building blocks of a Multilayer Perceptron. These entities loosely resemble their analogous biological counterpart: each unit receives a certain amount of input signals from other units, adds a custom bias term, performs a weighted sum and applies a nonlinear transformation in order to produce an output signal. 

Among the several Neural Network architectures that have been developed throughout the years, a particular class is that of Convolutional Neural Networks (CNNs). CNNs proved to be extremely powerful if applied to processes like image and video recognition and natural language processing. 

Mathematically, the bidimensional discrete convolution between two functions $F$ and $G$ is given by the following expression:
\begin{equation}
(F \circledast G) \ (i,j) = \sum_{m,n = -\infty}^{\infty} F(m,n) G (i-m,j-n).
\end{equation}

$F$ is referred to as the input function (a bidimensional grid-like object) while $G$ is called \textit{kernel function}. $G$ is much smaller than $F$.

In the vast majority of CNNs \cite{bengio, imagenet, flex_cnn} a convolutional layer does not contain only the convolution operation: in fact, it is followed by an activation layer and, usually, by a pooling layer. The activation layer transforms the feature map through the application of a nonlinear function \cite{bengio}. Pooling layers downscale the input data: given the output of the activation layer at a certain location, a pooling operation performs a summary statistic (average, maximum \cite{zhou}) on its neighbours that replaces the original value.

Common CNNs consist of a sequence of convolutional layers followed by a number of \textit{fully connected} (\textit{dense}) layers placed before the output. This is the network architecture of choice for this work, as detailed later on.

\subsection{Elastic Network Models}

Classical Molecular Dynamics \cite{md_general_method, md_sim_biomol}, by which Newton's equations of motion are numerically integrated, is the most effective and widespread  method used to investigate {\it in silico} the equilibrium properties and the dynamics of a (biological) molecule. Despite the recent dramatic gains in computational efficiency \cite{md_gpu, md_gpu_2, md_mbpot_gpu, md_anton}, many biological phenomena cannot be investigated with atomically detailed models: this is a particularly limiting problem if the system size exceeds few tens of millions of atoms or if the relevant biological processes occur over long timescales (typically larger than hundreds of microseconds). Furthermore, it is important to underline that highly detailed atomistic MD simulations generate an enormous amount of data, which is often difficult to store and post-process and, sometimes, simply not needed. 

MD simulations rely on sophisticated semi-empirical potentials that depend on a large number of parameters and reference properties; however, in a seminal paper Tirion \cite{tirion} showed that, in several cases, it is possible to replace the atomistic potential with a much simpler, single-parameter harmonic spring:
\begin{equation}\label{eq:ENM1}
E_{P} = \sum_{(i,j)} E(\bm{r}_{i} , \bm{r}_{j}) = \sum_{(i,j)} \frac{C}{2} \ \left( |\bm{r}_{i,j}| - |\bm{r}_{i,j}^{0}| \right)^2
\end{equation}
where the parentheses in the summation $(i,j)$ indicate that the sum is restricted to those atom pairs whose distance $|\bm{r}_{i,j}| = |\bm{r}_{i} - \bm{r}_{j}|$ is lower than a cutoff radius. 

This functional form of the potential is extremely simplistic, as 3-body terms are not even taken into consideration. Nevertheless, it can capture the collective, low energy vibrations of proteins. The slowest modes of vibration involve several atoms and interatomic interactions, whose sum approaches a universal form governed by the central limit theorem . For slow, collective modes, the details on the form of the pair potentials can be neglected \cite{tirion}, and if one is only interested in analyzing these modes (which usually dictate the function-oriented dynamics of the molecule) a single-parameter harmonic description can provide accurate predictions.

The potential energy in Eq. \ref{eq:ENM1} gives rise to the following Hessian matrix:
\begin{eqnarray}\label{eq:ENM2}
M_{ij,\mu\nu} &=& \frac{\partial^{2} V}{\partial x_{i, \mu} \partial x_{j,\nu} }\\ \nonumber
&=& - k_{ANM} \frac{(x_{i,\mu} - x_{j,\mu})(x_{i,\nu} - x_{j,\nu})}{|\bm{r}_{(i,j)}^{0}|^2}
\end{eqnarray}
where $\bm{x}_{i} =\bm{r}_{i} - \bm{r}_{i}^{0}$ and $\mu$ and $\nu$ are Cartesian components. Models described solely by the Hessian matrix in Eq. \ref{eq:ENM2} are called anisotropic elastic network models, or ANMs. The advantage of a quadratic approximation to Eq. \ref{eq:ENM1} is that the normal modes of vibration can be straightforwardly obtained through the inversion of the Hessian.

As anticipated, ENMs can be employed in contexts other than the analysis of vibrational spectra. In fact, it is possible to associate the harmonic force field of these models with simplified representations of the structure, that is, coarse-grained models. Coarse-graining can be defined as the process of reducing the accuracy and resolution of the representation of a system, describing it in terms of fewer collective degrees of freedom and effective interactions. The former are usually defined lumping together a relatively large number of atoms ($2-3$ to tens) into a single bead; the latter, on the other hand, are parametrised making use of one of the many available strategies \cite{Takada2012,Noid2013,Saunders2013,Potestio2014}, which in general aim at reproducing the multi-body potential of mean force of the system. Coarse-grained ENMs are typically constructed retaining only the $C_\alpha$ atom of the backbone, and placing a harmonic spring between pairs of atoms whose distance in the native conformation lies within a given interaction cutoff. More refined models employ also the $C_\beta$ carbon atom --or an equivalent one-- which explicitly accounts for the amino acid side chain. It is important to underline that the spring potentials employed in coarse-grained ENMs are a proxy for a thermal average of true all-atoms interactions over all conformations compatible with a given coarse-grained configuration; hence, they consist in free energies rather than potential energies, as it is usually the case in the context of coarse-graining.

Studies making use of all-atom or coarse-grained ENMs proved to be particularly effective in the modelling and prediction of low energy conformational fluctuations, corresponding to the most collective normal modes. These results are often in agreement with the ones produced using all-atoms MD simulations with a standard semi-empirical force field. Among the most notable structures they have been applied to we point out RNA Polymerase II \cite{rnapoly}, Virus Capsids \cite{virus}, Transmembrane Channels \cite{tch} and the whole ribosome \cite{ribosome}.

The $\beta$-Gaussian Model ($\beta$-GM \cite{betagm}) is a particular flavour of coarse-grained ENM in which the description of protein fluctuations is improved through the introduction of effective $C_{\beta}$ centroids (with the exception of Glycine residues, whose side chain is made up by a single hydrogen atom). The $\beta$-GM model is defined on a coarse-grained protein structure, thus providing a simplified description of the system's fluctuations in a local free energy minimum, centred on a reference structure typically chosen to be the native, crystallographic conformation. The introduction of $C_{\beta}$ atoms in the model results in a Hamiltonian that is considerably more complex than the one relative to a chain of $C_{\alpha}$ units, whose general form is given by:

\begin{equation}
\begin{split}
\ch{} = \frac{1}{2} \sum_{i,j} \bm{x}^{C_\alpha}_{i} \bm{M}_{i,j}^{C_\alpha-C_\alpha}\bm{x}^{C_\alpha}_{j}\\
+ \sum_{i,j} \bm{x}^{C_\alpha}_{i} \bm{M}_{i,j}^{C_\alpha-C_\beta}\bm{x}^{C_\beta}_{j}\\
+ \frac{1}{2} \sum_{i,j} \bm{x}^{C_\beta}_{i} \bm{M}_{i,j}^{C_\beta-C_\beta}\bm{x}^{C_\beta}_{j}
\end{split}
\end{equation}
where $M$ is an interaction matrix. The first term of the Hamiltonian represents the interactions between $C_{\alpha}$ atoms, be they bonded links between consecutive $C_{\alpha}$'s along the peptide chain or simply those belonging to close-by residues in contact in the native conformation. The second term accounts for interactions between $C_{\alpha}$'s and $C_{\beta}$'s; lastly, the third term includes interactions existing solely among $C_{\beta}$'s.

In the $\beta$-GM framework, the positions of $C_{\beta}$'s are assigned using a simplified version of  the Park-Levitt procedure \cite{park-levitt}, in which the  $C_{\beta}$ centroids are placed in the plane specified by the local $C_{\alpha}$ trace. This assumption allows one to compute the $C_{\beta}$ coordinates of atom $i$ using only the positions of $C_{\alpha}$'s $i-1$, $i$ and $i+1$, giving rise to a Hamiltonian that is quadratic in the $C_{\alpha}$ deviations but with a different coupling matrix. This model has the very same computational cost of less accurate, $C_{\alpha}$--only anisotropic models, but it is able to capture in a more accurate way the low-energy macromolecular fluctuations. In this way the deviations of $C_{\beta}$ atoms of all amino acids (excluding Glycine and the terminal residues) are parametrised using the $C_{\alpha}$ Cartesian coordinates, leading to a new Hamiltonian of the form:
\begin{equation}\label{hbgm}
\tilde{\ch{}} = \frac{1}{2}  \sum_{i,j} \bm{x}^{C_\alpha}_{i} \tilde{\bm{M}}_{i,j}^{C_\alpha-C_\alpha}\bm{x}^{C_\alpha}_{j}
\end{equation}

In the present work we have been consistent with the model as described in the original paper \cite{betagm}, and used the same parameters present therein. In particular, the cutoff radius $R_{c}$ has been set to $7.5$ \AA.

\subsection{Construction of the protein dataset}

\begin{figure}
\centering
\includegraphics[width=\columnwidth]{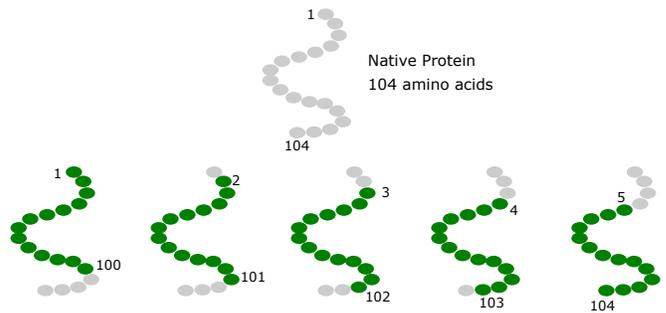}
\caption{\label{fig:decoys} Schematic illustration of the procedure followed to construct the 100-amino acids long decoys. Proteins whose sequence is longer than 100 residues are cut in 100-residues long sub-sequences sliding a window of this length along the main chain. A protein of length $100+N$ amino acids produces $N+1$ decoys.}
\end{figure}

In the previous section we described the exactly solvable algorithmic procedure through which one can compute eigenvalues and eigenvectors associated to the local fluctuation dynamics of the $\beta$-GM coarse-grained protein model. As anticipated in the introduction, the scope of our work consists in building a DL architecture (CNN) capable of predicting the lowest ten of these eigenvalues. In order to do so we have to first train and subsequently validate this CNN approach. We constructed two separate groups of protein structures, downloaded from the Protein Data Bank (PDB), to be used as Training and Evaluation Sets, respectively. The Evaluation Set contains protein structures with a single chain and 100 $C_{\alpha}$ atoms; for the Training Set we considered chains with a length between 101 and 110 monomers that have been processed to construct $N+1$ \textit{decoys} for each protein of length $100+N$. In this specific context, by decoy we indicate protein-like chains or sub-chains that preserve the vast majority of typical structural properties of real, ``full'' proteins \cite{decoys_roitberg}. Figure \ref{fig:decoys} illustrates the procedure followed to produce such decoys.

Through this simple process we obtain an Evaluation Set of 146 real proteins with 100 amino acids and a Training Set of 10728 decoys of the same length. It is important to notice that the $\beta$-GM spectrum is invariant with respect to the \emph{orientation} of the sequence, namely we can easily double both datasets including the reversed structures.

\begin{figure}
\centering
\includegraphics[width=\columnwidth]{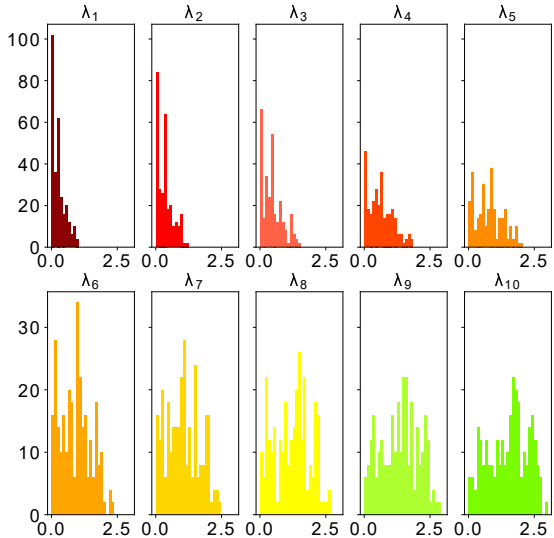}
\caption{\label{fig:spectra}Distribution of the lowest 10 eigenvalues of the spectrum of all proteins in the Evaluation Set, computed by means of the $\beta$-GM.}
\end{figure}

Dealing with proteins and  biologically relevant decoys we encounter a wide variety of structures. They are extremely heterogeneous in terms of radius of gyration and their spectra show high variability. Fig. \ref{fig:spectra} shows the distribution of the ten lowest eigenvalues in the Validation Set. It can be seen that the $\lambda_{i}$'s become more broadly distributed with increasing $i$, a feature that has been previously observed in the context of ensemble analysis of ENM spectra \cite{potestio2010prl}.

\begin{figure}
\centering
\includegraphics[width=\columnwidth]{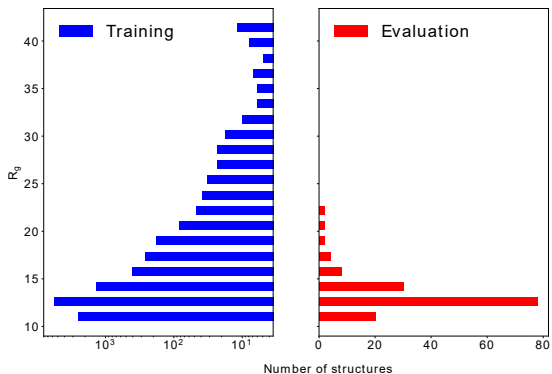}
\caption{\label{fig:bgm_gyra}Distribution of the radius of gyration over Training set (left) and Evaluation set (right).}
\end{figure}

Fig. \ref{fig:bgm_gyra} shows a histogram of the globularity of the samples in the available datasets in terms of radius of gyration. In the Training Set there are several structures that are highly non-globular, but the vast majority has a gyration radius comparable to the values present in the Evaluation Set.

\subsection{Molecular descriptor}

A crucial step in the construction of a DL-based pipeline to analyze and process a given molecular structure is the identification of an appropriate \textit{molecular descriptor}. From Eq. \ref{hbgm} we can see that, within the $\beta$-GM framework, the Hamiltonian of the system depends only on the positions of the $C_{\alpha}$ atoms. Hence, our molecular descriptor will take as input only the Cartesian coordinates of these atoms. However, for CNN applications we cannot simply characterise the biomolecule in terms of Cartesian coordinates, since these are not invariant with respect to rotations and translations of the system, an important requirement a molecular descriptor has to fulfill.

A prominent example of molecular descriptor is given by Behler and Parrinello's \textit{symmetry functions} \cite{bp,behler}. These functions describe the local environment of each atom in a molecular system, while satisfying the invariance requirement. Among the parameters that are defined in the calculation of these quantities, the most relevant one from a conceptual point of view is the cutoff radius: interatomic distances larger than this value yield zero contribution to these descriptors. Symmetry functions have been mainly employed in order to provide accurate potential energy surfaces \cite{behler,bp} and to detect local atomic arrangements in liquids \cite{dellago}. Although these descriptors proved to be extremely successful, our learning task concerns the prediction of a property that is (at least in general) intrinsically global, hence we need a function that is able to encode all the interactions between the atoms that constitute the system. Therefore we decided to characterise the proteins under examination in terms of the Coulomb Matrix, a very general and global molecular descriptor that is rotation-translation invariant. This is defined as:

\begin{equation}\label{eq:cm}
    M_{IJ} = 
    \begin{cases}
      0 & \text{if}\ I=J \\
      \frac{1}{|R_{I} - R_{J}|} & \text{otherwise}
    \end{cases}
\end{equation}
where $I$ and $J$ are atomic indices. It is important to remark that in our work we did not consider properties related to the charge, so we set all the diagonal elements in Eq. \ref{eq:cm} to zero.

\subsection{Architecture}

In the previous sections we defined all the elements of our learning problem, namely the chosen molecular descriptor, the desired output, and the algorithmic procedure used to produce it.  We now illustrate the architecture of the network employed.

The motivations behind the choice of a CNN architecture to address the problem at hand are essentially three. First, \textit{Parameter Sharing} allows one to keep the total number of parameters to be \textit{learned} relatively low. If we used an ANN, which has \emph{tied} weights, we would have obtained a much higher total number of learnable parameters. Second, CNNs are particularly suited to deal with grid-like input data, such as Coulomb matrices. Third, no data preprocessing is required.

\begin{figure*}
\centering
\includegraphics[width=0.9\linewidth]{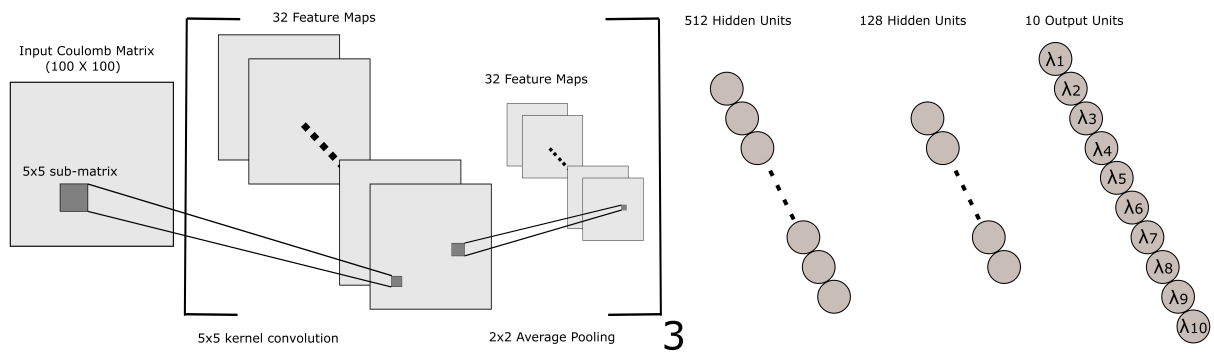}
  \caption{The CNN architecture employed in the present work. A convolutional layer is drawn between squared brackets to emphasize that it is iteratively repeated for three times. For the sake of clarity, connections between neurons in the fully connected layers are not represented.}
  \label{fig:cnn_full}
\end{figure*}

Here we used a CNN composed by three convolutional layers and two fully connected layers. Each convolutional layer is made by a convolution operation followed by an average pooling layer. While the latter acts on regions of amplitude $2 \times 2$, the former is realised with the use of 32 kernel functions, each of which is a $5 \times 5$ matrix whose elements represent the learnable parameters (weights). The dense layers consist of 512 and 128 units, respectively. There are ten output units, each of which corresponds to one non-zero eigenvalue of the $\beta$-GM spectrum. The network structure is sketched in Fig. \ref{fig:cnn_full}. Three \textit{dropout} \cite{dropout} layers have been included in the network before, between, and after the fully connected layers. Dropout is a \textit{regularization} technique that drops a certain ratio (25\% in our case) of the input units of a layer at each step of the training process. This technique significantly prevents the risk of overfitting the training set \cite{dropout_2}.

The network is developed using Keras \cite{keras} with Tensorflow \cite{tensorflow} backend. The optimizer is adagrad \cite{adagrad} with learning rate = 0.008. The batch size is 400 and the number of epochs is bound to 100. For what concerns the loss function we decided to employ the mean absolute percentage error (MAPE):

\begin{equation}\label{eq_mape}
MAPE =  \frac{1}{N} \sum_{j=1}^{N} \frac{1}{10} \sum_{i = 1}^{10} \frac{\abs{\lambda_{i}^j - \hat{\lambda}_{i}^j}}{\abs{\hat{\lambda}_{i}^j}}
\end{equation}
where $N$ is the batch size while $\hat{\lambda}_{i}^j$ and $\lambda_{i}^j$ represent true and predicted eigenvalues, respectively. Recent work by de Myttenaere {\it et al.} \cite{demytt1} proved generic consistency results for this loss function.

In order to quantitatively assess the effectiveness of the network, we analysed the CNN-predicted eigenvalues through a cross-validation procedure. This is a common strategy to evaluate the performances of a learning algorithm and its ability to generalise to an unknown and independent dataset. The idea behind this technique is the repetition of training and testing processes on different subsets of the full training set. $k$-fold cross-validation is the most known example of this procedure: the full training set is split into $k$ different \emph{folds}; for each of these subsets the algorithm is trained over the other $k-1$ folds and is tested against the unknown samples present in the $k$-th fold. In this work we have made use of the Deep Analysis Protocol (DAP) for cross-validation. This protocol has been extensively employed in many machine learning challenges applied to biological data \cite{maqc,seqc}, inducing an effective massive replication of data. In this work we performed a $10 \times 5$ cross-validation, namely a 5-fold cross-validation performed ten times, with ten different random seeds for the network. These are the same seeds that have been used during the process of training on the full dataset.

\section{Results and discussion}

Before discussing the results we deem it useful to highlight a few crucial aspects of the purpose of our work. In essence, the problem we tackle here can be seen as a spectral inversion by means of a CNN. In the literature there are previous examples \cite{ZHANG_YI_2003, FINOL_2018} of machine learning-based approaches to extract the eigenvalues of a matrix using mainly recurrent neural networks. However, our work focuses on an intrinsically different goal: first, we did not consider the actual interaction matrix of proteins as input data, rather the far more general distance matrix; second, our scope is to provide a preliminary example of how to employ DL-based algorithms to extract non-trivial, global structural properties of proteins. Our choice to make use of ENMs relies both on their simplicity and low computational requirements, which allowed us to quickly validate the performance of the CNN.

\begin{figure}
    \centering
    \includegraphics[width=\columnwidth]{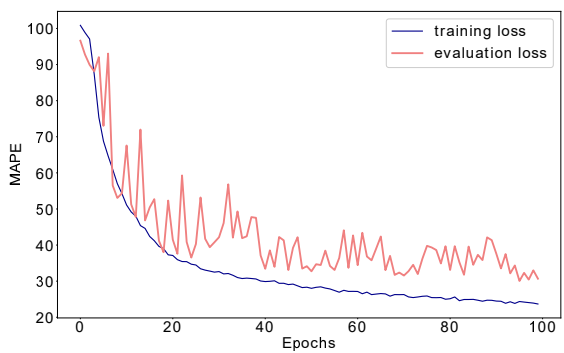}
    \caption{\label{fig:bgm_losses}Example of the behaviour of losses against the number of epochs. Both training loss and evaluation loss refer to the MAPE (Eq. \ref{eq_mape}), computed on the samples in the Training and Evaluation set respectively.}
\end{figure}

This validation was carried out through the application of the DAP to our multitask regression problem. In the several 5-fold cross-validation processes, the independent folds were built so that a structure and its \emph{reversed} counterpart were included in the same fold. On the other hand decoys coming from the same protein were allowed to be part of different folds. This results in folds that are not completely independent. In Fig. \ref{fig:bgm_losses} we can see an example of the behaviour of Training and Evaluation losses during the Training Process on the full Training Set. The losses have a quite steep decrease during the early stages of the training, where they are almost \emph{coupled}. After a few ($\sim20$) epochs the loss on the Evaluation Set starts oscillating, but it keeps decreasing. These non-negligible oscillations are due partly to the small size of the Evaluation Set, and partly to the relative lack of robustness of MAPE to small fluctuations.

\begin{figure}
  \centering
  \includegraphics[width=\columnwidth]{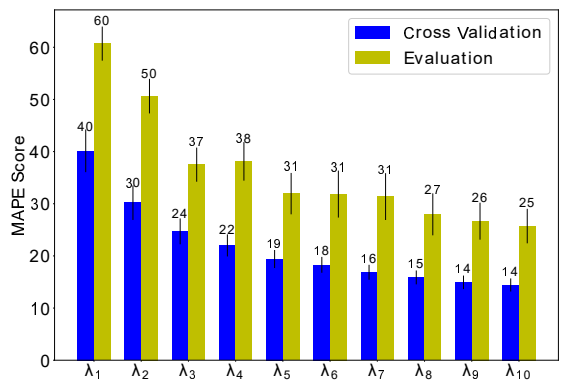}
  \caption{MAPE scores as obtained for simple evaluation (green bars) and cross validation (blue bars), broken down to each eigenvalue separately. The specific MAPE value of each bar is indicated above the latter.}
  \label{fig:cv_ev}
\end{figure}

\begin{figure}
  \centering
  \includegraphics[width=\columnwidth]{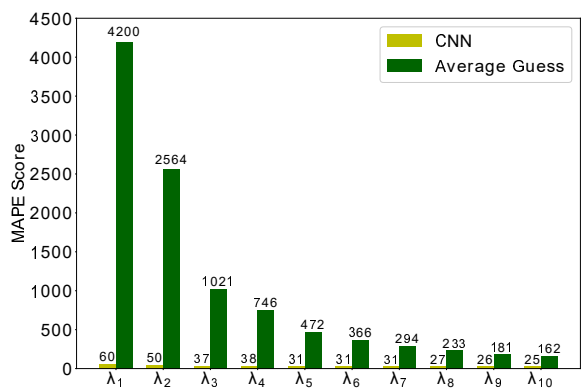}
  \caption{Comparison of the MAPE score as obtained from the CNN trained in the present work (light green bars) and a reference, maximally uninformative random model (dark green bars). The values attained by each model for each eigenvalue are indicated above the bar.}
  \label{fig:ave_guess}
\end{figure}

The result achieved for each eigenvalue in cross-validation and Evaluation are reported in Fig. \ref{fig:cv_ev}. Results in cross-validation are more accurate than the others: this is reasonable since we decided to include decoys generated from the same protein in different folds. MAPE is a relative performance measure; hence, in order to further assess the validity of our predictions, they have to be compared  to the ones given by a non-informative model. In Fig. \ref{fig:ave_guess} we can see a comparison between our results on the Evaluation Set and a non-informative model that always predicts the average value of each eigenvalue in the Training Set. In our case we can see that the predictions are considerably more accurate than the ones produced by this non-informative model.

\begin{figure*}
	\centering
	\begin{subfigure}{.5\textwidth}
		%\centering
		\includegraphics[width=\columnwidth]{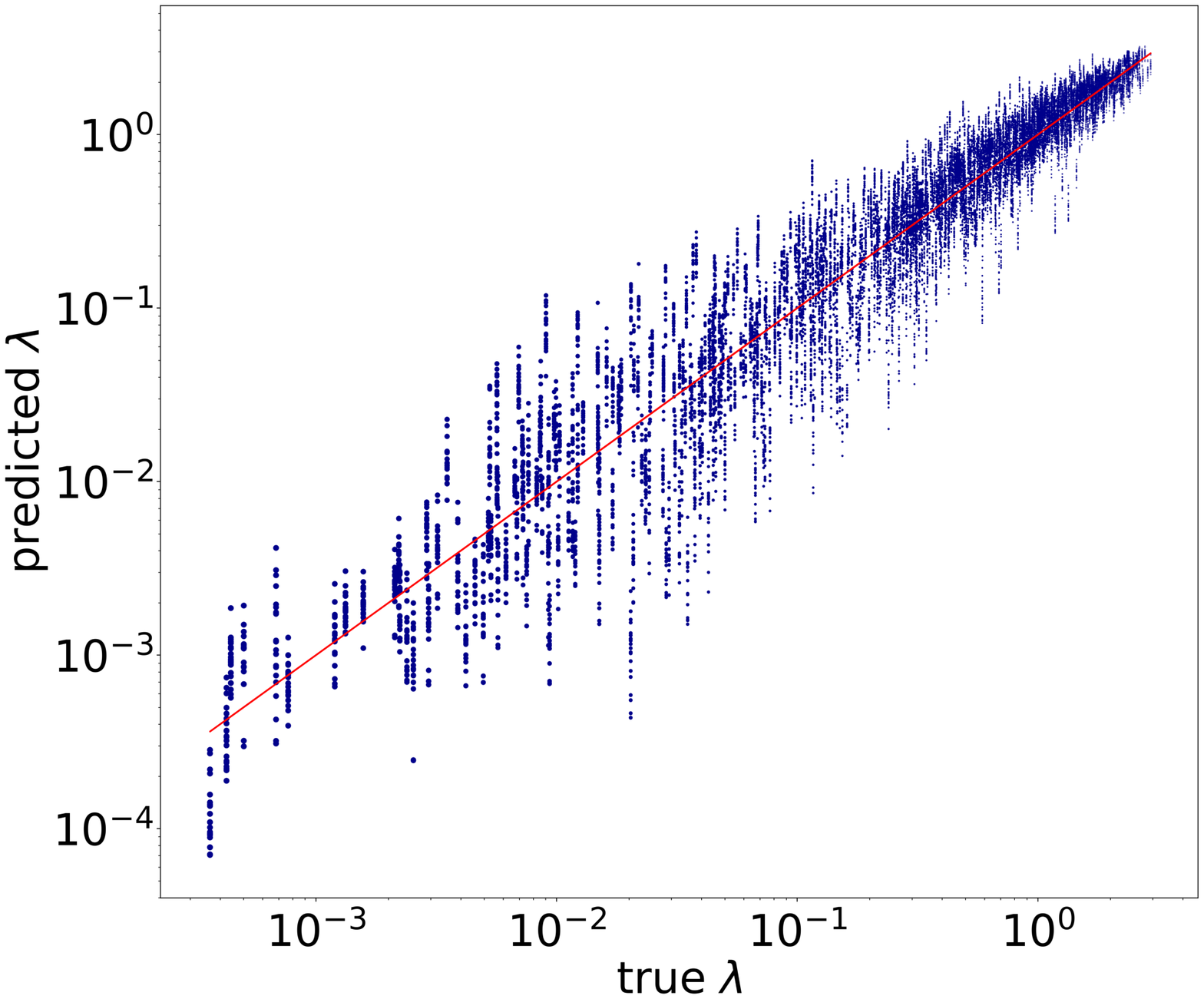}
		%\vspace{2mm}
		%\caption{}
		\label{fig:scatter1}
	\end{subfigure}%
	\begin{subfigure}{.5\textwidth}
		%\centering
		\includegraphics[width=\columnwidth]{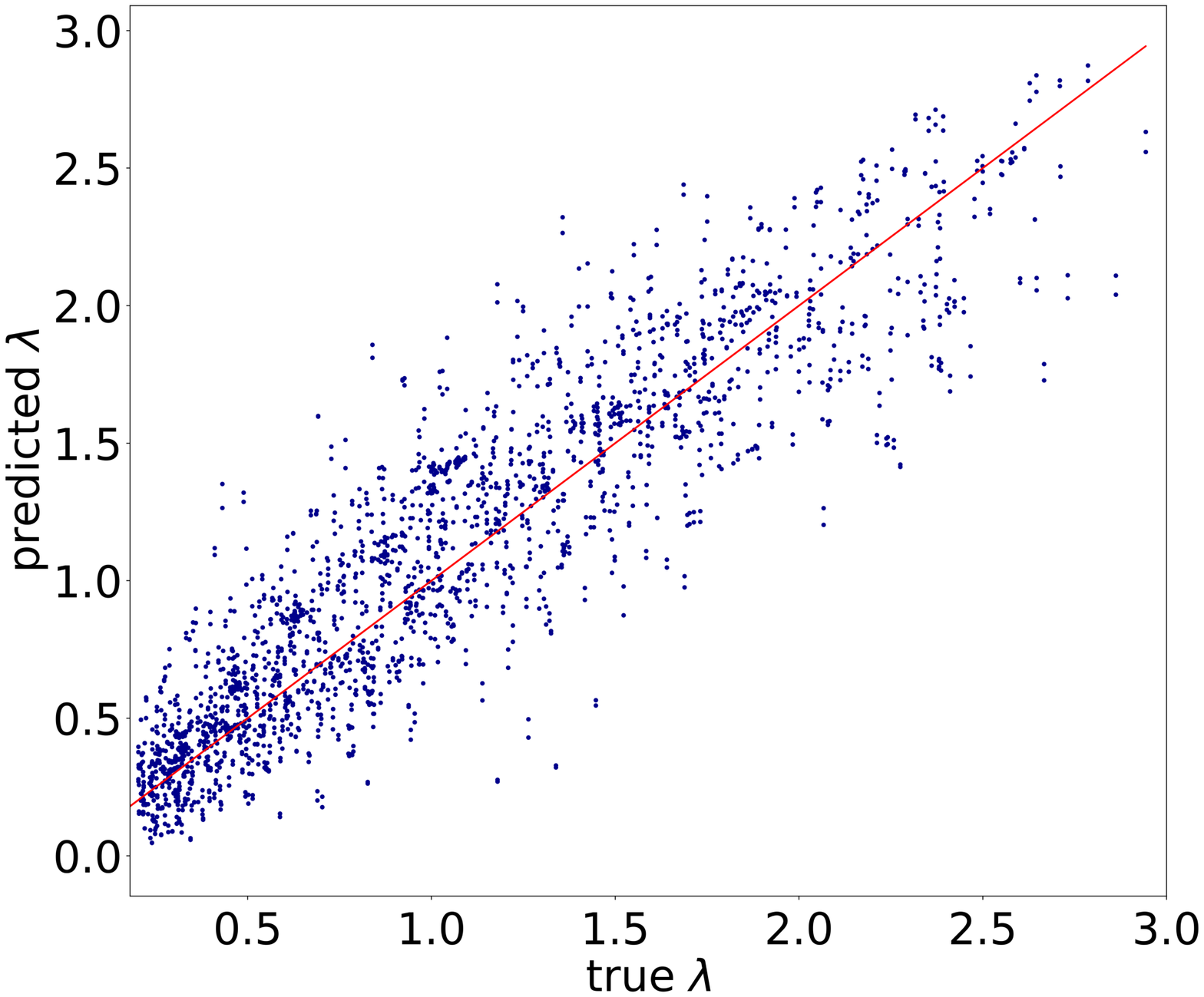}
		%\caption{}
		\label{fig:scatter2}
	\end{subfigure}
	\caption{\label{fig:scatter} Scatter plot of the eigenvalues predicted by the CNN against the exact ones computed by the $\beta$-GM. All lowest ten non-zero eigenvalues of each protein in the validation set (both ``direct'' and ``reverse'' orientation of the chain) computed by each of the 100 CNN instances are shown. Right panel: the data are shown in log-log scale. Left panel: data are reported in linear scale, however only those eigenvalues having $\lambda > 0.2$ can be seen.}
\end{figure*}

Fig. \ref{fig:scatter} shows the scatter plot of all the eigenvalues in the Evaluation Set plotted against their predicted values. Since we ran ten different experiments we had 20 predictions associated to a single real eigenvalue (sequence-reversed structures share the $\lambda$'s of the original structures). The almost linear behaviour suggests that the learning model is able to detect with good precision all the eigenvalues even if they range over several orders of magnitude.

The accuracy of the CNN-based approach leaves room for improvement, e.g. through an increased size of the training dataset, a more refined cost function, different parameters and structure of the network etc. However, there is a limitation in the proposed algorithm which is more fundamental than those mentioned, namely the fixed size of the input structures. In fact, the far-reaching objective of employing deep learning approaches for structural analysis of proteins would be severely limited if only structures with a given number of amino acids could be analysed. A mitigation of this issue comes from the nature of the problem under examination and, as it will be illustrated hereafter, opens a novel scenario for the usage of CNNs in the present context.

A few words are in place regarding the computational cost of the CNN in comparison with the exact algorithmic procedure. The time required by the network-based approach to process the entire training set and predict the corresponding eigenvalues is shorter than 5 minutes on a single-core CPU, while the application of the $\beta$-GM to the same dataset requires 25 minutes on the same platform. It is evident that the amount of information provided by the two approaches, as well as the relative accuracy, are not comparable: while the $\beta$-GM produces the full, {\it exact} spectrum for each molecule (eigenvalues as well as eigenvectors), the CNN can only provide an estimate of the ten lowest eigenvalues. Nonetheless, even though the computational gain obtained already in this case is substantial, one has to bear in mind that the ENM is here taken as reference algorithm precisely because of its velocity and accuracy; on the contrary, we envision applications involving much more time-consuming procedures, e.g. the optimization of complex cost functions \cite{digginsJCTC2018}, for which the speedup can be substantial.

The defining property of low-energy modes of fluctuations is their collective character, which manifests itself in the fact that several residues are displaced in the same direction, with no or very little strain among them. This characteristic lies at the foundation of coarse-graining approaches which aim at identifying large groups of residues behaving as quasi-rigid units \cite{Hinsen1998,Gohlke2006,ZHANG_BJ_2008,ZHANG_BJ_2009,Potestio2009,aleksiev2009,ZHANG_JCTC_2010,Sinitskiy2012,Polles2013}. It is thus the case that the residues which determine the low energy eigenvalues in ENMs are those few whose distances vary the most, that is, hinge residues. Consequently, it is reasonable to expect that the elimination of a few $C_\alpha$'s from the model would not too drastically impact the value of the computed spectra.

\begin{figure*}
\centering
\includegraphics[width=2\columnwidth]{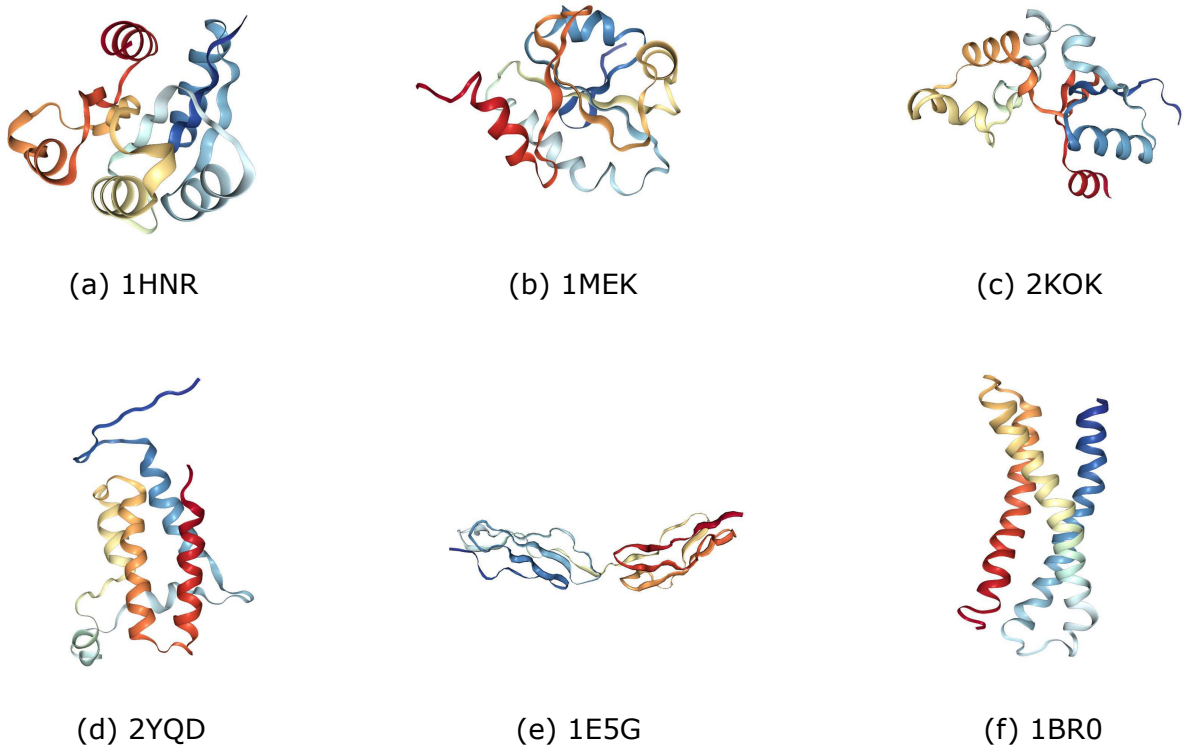}
\caption{Structures of the six proteins with 120 amino acids employed to test the behavior of the CNN on molecules larger than those employed for the training. The PDB codes are indicated. The graphical representation has been done using the online tool available on the website of the Protein Data Bank \cite{PDB_NAR2000}.}
\label{fig:6prots}
\end{figure*}

\begin{figure}
  \centering
  \includegraphics[width=1\columnwidth]{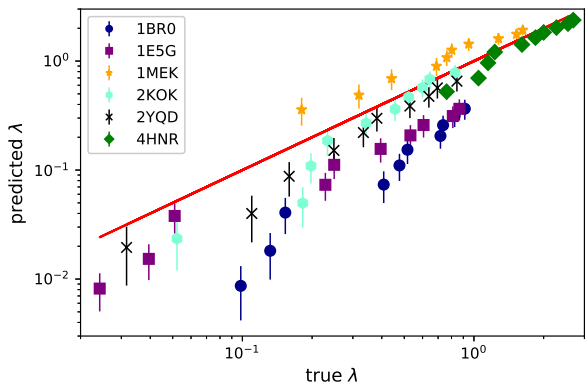}
  \caption{Predicted eigenvalues of the six structures with 120 amino acids against their real value. Each point is the average over 100 random coarse-graining procedures and 10 networks, error bars indicate the standard deviation. The red line is a guide to the eye.}
  \label{fig:random_cg_errorbars}
\end{figure}

To provide quantitative concreteness to these hypotheses, we fed the CNN, trained to intake 100-residues-long structures, with six proteins of 120 amino acids, 20 of which have been randomly pruned. In Fig. \ref{fig:6prots} we show the structure of the selected molecules, which have been chosen from the PDB so as to have some degree of structural variability. These proteins range from very globular (4HNR) to fairly elongated (1BR0) ones, up to a case where a hinge is evident and identifiable already by visual inspection (1E5G). For each of these six molecules we realised 100 different coarse-grained structures having only 100 amino acids by randomly removing 20 of them. The model set of each protein has been fed to 10 networks, differing only for the initial guess of the hyperparameters. In Fig. \ref{fig:random_cg_errorbars} we report the average of the first 10 eigenvalues of each of the 6 proteins, averaged over the 100 randomly pruned structures and the 10 CNN instances. These eigenvalues are plotted against the value computed by means of the $\beta$-GM.

A few observations are in order. In one case, namely 4HNR, there is a perfect overlap between the predictions on the randomly coarse-grained structures and the actual values, with an overall average MAPE equal to 15.8. This molecule is highly globular, which also reflects in the large absolute value of the eigenvalues; hence, it seems that the removal of a relevant fraction of amino acids does not affect the precision of the CNN model. Eigenvalues associated to 2KOK, 2YQD, and 1MEK were predicted with reasonable accuracy, the overall average MAPE being 30.7, 35.5, and 48.3, respectively. These proteins share a medium degree of globularity. The most important deviations from the real eigenvalues appear for 1BR0 and 1E5G, with largely underestimated values; for these molecules, the overall average MAPE equals 73.7 and 58.3, respectively. However, these proteins are at the other extreme of the ``globularity spectrum'' with respect to the first, very compact 4HNR. In fact, 1BR0 is a bundle of 3 quasi-parallel alpha-helices, while 1E5G consists of two identical, independent domains separated by a few interface residues. It is reasonable to expect that in the first case no well-defined hinge region exists, rather each part of the molecule takes part in the low-energy deformation. A random removal of residues thus has an appreciable impact in the calculation of the energetic cost associated with collective motions.

\begin{figure}
  \centering
  \includegraphics[width=\columnwidth]{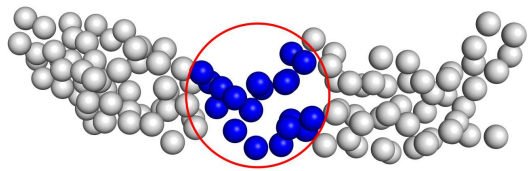}
  \caption{Schematics of the procedure to perform a restrained random removal of the exceeding 20 amino acids from protein 1E5G. Atoms to be eliminated can be selected only outside of the sphere centered on the molecule's hinge and having a 1 nm radius.}
  \label{fig:sphere_proc}
\end{figure}

\begin{figure}
  \centering
  \includegraphics[width=\columnwidth]{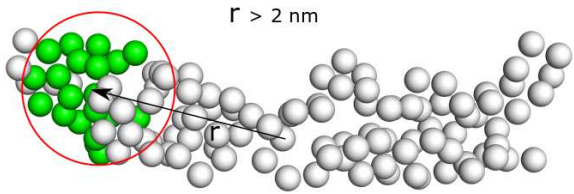}
  \caption{Schematics of the procedure to test whether the improved MAPE values obtained excluding a localized group of residues from removal does not depend on their location. As described in Fig. \ref{fig:sphere_proc}, the 20 residues to remove can be randomly selected only outside of a sphere of 1 nm radius. The center of the sphere, however, cannot be localized closer than 2 nm to the point employed for the previous analysis, namely the molecule's sequence center (i.e. the mechanical hinge). 10 different positions for the sphere are randomly identified, and for each of them 10 different models where 20 residues have been randomly removed have been constructed.}
  \label{fig:random_sphere_proc}
\end{figure}

\begin{figure}
  \centering
  \includegraphics[width=\columnwidth]{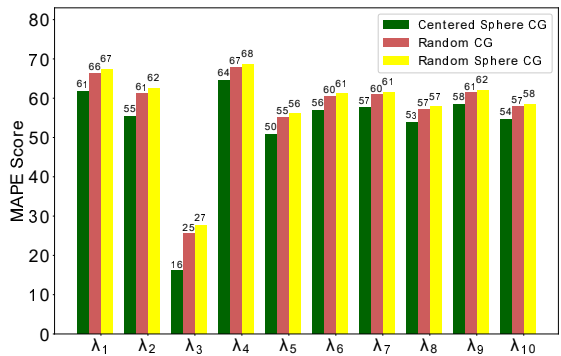}
  \caption{1E5G: MAPE of predictions on proteins subject to different procedures for the removal of the 20 exceeding residues. \textit{Random CG} stands for randomly coarse-grained structures, \textit{Centred Sphere CG} refers to the procedure in which central atoms are not removed; and \textit{Random Sphere CG} describes structures in which the sphere has been placed in regions other than the interface between the domains.}
  \label{fig:sphere}
\end{figure}

For 1E5G the mechanism is different. This protein possesses a short linker and a relatively small interface connecting two lobes, thus suggesting a rather decoupled dynamics between them. That this is likely the case is made evident by the fact that this molecule features the lowest lowest-energy (sic) eigenvalue among those under examination. Hence, the removal of some residues from those constituting the hinge between the two domains substantially affects the result. In order to verify this hypothesis, we have repeated the CNN-based calculation for 1E5G on a set of 100 pruned structures, which have been obtained randomly removing 20 amino acids from the crystallographic conformation with the restraint that no residue lying within a 10\AA\ cutoff from the center of the linker could be eliminated, as illustrated in Fig. \ref{fig:sphere_proc}. The effect of this simple criterion can be seen on the data reported in Fig. \ref{fig:sphere}, which show a small yet appreciable and, most importantly, systematic improvement of the MAPE score when the hinge residues are not selected for removal ({\it centred sphere CG}) with respect to the completely random selection case ({\it random CG}).

Finally, to rule out the possibility that this improvement depends on the exclusion of a localized group of CG sites {\it per se} rather than their particular location, we have performed a further test. Specifically, we have constructed ten different CG model types in which the 20 exceeding residues have been eliminated outside of a sphere of radius 10\AA\ whose center is located on a randomly chosen position of the protein at least 20\AA\ away from the hinge centre. In plain English, we have performed the same calculation as of the centred sphere CG ten times, with spheres centred so as to avoid overlap with the one placed in the protein hinge. For each model type --i.e. for each location of the exclusion sphere-- ten randomized CG models have been produced, and their eigenvalues averaged over specific coarse-graining realization and CNN model.

The result, also visible in Fig. \ref{fig:sphere}, shows an {\it increase} of the MAPE score with respect to the random case, that is, the prediction of the CNN worsens with respect to a model where the 20 removed residues have been randomly chosen throughout the structure. This observation consolidates the hypothesis that the network is capable of predicting with sufficient accuracy the low-energy eigenvalues of a protein larger than those it has been trained upon, provided that the exceeding number of sites has been removed; furthermore, and quite intuitively, the prediction improves if the removed residues do not belong to mechanically relevant parts of the molecule such as motion hinges.

\section{Conclusions}

The aim of computer-aided modelling of biomolecular systems it to achieve deeper, mechanistic understanding of their function and properties at a level that cannot be accessed by means of experimental or purely theoretical (i.e. mathematical) methods. This approach indeed plays on two sides of a medal: on the one hand, it provides a detailed picture of biological processes at the molecular level, thus enabling the confirmation or falsification of hypotheses and the formulation of theories and models of the finest mechanisms of living matter; on the other hand, it serves as a validation of the currently available representations of the fundamental constituents of cells, ranging from single atoms to entire tissues and organs. Such a workflow is largely {\it algorithmic and deterministic}, in the sense that it relies on well-defined procedures each step of which is known and understood. An exemplary instrument in this sense is molecular dynamics.

The alternative strategy, which is gaining further and further attention and interest (as well as success), is machine learning, and deep neural networks in particular. These computational methods have proven extremely effective in performing those tasks which cannot be easily formulated in a classically {\it algorithmic} manner, rather they have a fuzzier, more probabilistic character. Nonetheless, a steadily growing level of quantitative accuracy is being reached by deep learning techniques.

The complementary nature of the two aforementioned approaches is not only extremely appealing, but also potentially very powerful, as it is demonstrated for example in the field of bioinformatics, where (big) data processing moves on both tracks simultaneously. In the present work we have made a first attempt to combine formal, algorithmic models with deep learning approaches in the context of protein modelling. In particular, it has been our goal to perform, by means of a convolutional neural network, the calculation of global properties of protein structures such as the lowest-energy eigenvalues of the most collective modes of fluctuations. The final aim cannot, of course, be that of trivially replacing the simple, extremely effective procedure represented by a matrix inversion by means of a CNN; rather, we explored the possibility of allowing a deep learning scheme to perform this task with sufficient accuracy as a first, necessary step towards more complex kinds of structural protein analyses. While the calculation of the lowest eigenvalues (as well as the rest of the whole spectrum) of an elastic network model is immediate and computationally inexpensive in terms of linear algebra, it is not given for granted that a CNN could do it as well. Furthermore, a crucial step in the usage of a CNN (or similar methods) is the pre-processing of the molecular structure in terms of appropriate input variables: the usage of the Coulomb matrix has proven to be a viable choice to this end.

A second, equally relevant outcome of this work is the extension of the network-based eigenvalue prediction network to proteins having a larger number of residues than those employed for the training. The construction of molecular descriptors flexible enough to process proteins of variable length is still an open issue; however, we have shown that the network --trained on 100-residue-long molecules-- can provide good estimates of the low-energy eigenvalues of proteins with 120 amino acids provided that the exceeding ones have been neglected. This positive result proves even more pleasant inasmuch as the agreement between predicted and real values varies depending on the specific choice of the removed amino acids, in such a way that mechanically relevant residues emerge as those whose removal determines a worsening of the prediction. The natural consequence of this observation is that, upon appropriate training, deep learning schemes could be employed in an effective manner not only to compute properties along the lines of reference algorithms, but also to extract biologically relevant features of a protein and to provide valuable indication on how to construct simplified, that is, coarse-grained models.

\begin{acknowledgements}
We thank Simone Orioli and Flavio Vella for a critical reading of the manuscript and insightful discussion. RP gratefully acknowledges the support provided by the Royal Society in the funding and organization of the Theo Murphy international scientific meeting {\it Multi-resolution simulations of intracellular processes} which took place on Sept. 24--25, 2018 at Chicheley Hall, Buckinghamshire (UK), during which inspiring and fruitful conversations took place. This project has received funding from the European Research Council (ERC) under the European Union's Horizon 2020 research and innovation programme (grant agreement No 758588).
\end{acknowledgements}

\bibliographystyle{ieeetr}
\bibliography{references}

\end{document}